\documentclass[5p,sort&compress,times,letterpaper]{elsarticle}
\usepackage{amsmath,amssymb}

\usepackage{bm}
\usepackage{graphicx}
\usepackage{color}

\usepackage{epsfig}
\usepackage{graphicx}
\usepackage{dcolumn}
\usepackage{bm}
\usepackage{multirow}
\usepackage{array}
\usepackage{slashed}
\usepackage{amsmath}
\usepackage[titletoc]{appendix}
\usepackage{color}
\usepackage{diagbox}

\begin{document}

\title{String effect on the relative time delay in the Kerr-Sen black hole}

\author[BSPU]{R.N. Izmailov}
\author[BSPU]{R.Kh. Karimov}
\author[BSU]{A.A. Potapov}
\author[BSPU,NBU]{K.K. Nandi}
\address[BSPU]{Zel'dovich International Center for Astrophysics, Bashkir State Pedagogical University, 3A, October Revolution Street, Ufa 450008, RB, Russia}
\address[BSU]{Department of Physics \& Astronomy, Bashkir State University, 47A, Lenin Street, Sterlitamak 453103, RB, Russia}
\address[NBU]{High Energy Cosmic Ray Research Center, University of North Bengal, Darjeeling 734 013, WB, India}

\date{\today}

\begin{abstract}
A well known solution of heterotic string theory is the spinning Kerr-Sen black hole (KSBH) characterized by a string parameter $\xi$. Kerr black hole is recovered at $\xi=0$. The purpose of this paper is to investigate the effect of $\xi$ on a new diagnostic of relative time delay (RTD) to see how the latter deviates from that in general relativity. Assuming KSBH as the spinning lens partner in PSR-BH binary systems, which provide the best laboratory for testing the time delay predictions, we study here the RTD up to third PPN order in $\left(1/r\right)$ in the thin-lens approximation. We work out a useful generalization of the RTD formulas applicable to the experimentally viable \textit{finite} distance lens scales, while\ terms higher than the zeroth order are shown to contain the effect of $\xi$. We shall also relate RTD to the observable image magnification factor determined by $\beta/\theta_{E}$, where $\beta$ is the angular separation between the source and the observer and $\theta_{E}$ is the "Einstein angle" determined by an "effective" non-aligned static lens equivalent to the original aligned spinning lens. Numerical estimates for two typical binary lens systems show $\mu$sec level delay at the zeroth order consistent with predictions in the literature. However, the string effect at higher orders is too tiny to be measurable even in the far future leading to the conclusion that the stringy and general relativity BHs are yet observationally indistinguishable.
\end{abstract}

\maketitle

\section{Introduction} \label{sec:intro}
A black hole solution of heterotic string theory in 4 dimensions is the asymptotically flat spinning Kerr-Sen black hole (KSBH) characterized by a string parameter $\xi=Q^{2}/2M$, where $Q$ is the dilatonic charge and $M$ is the mass \cite{sen:1992}. The KSBH received a lot of attention (see, e.g., \cite{bhadra:2002}) since its discovery by Sen. We call it here KSBH, since the solution is a string generalization of the Kerr BH \cite{kerr:1963} of general relativity theory, the latter BH being recovered at $\xi=0$. A natural enquiry then is how $\xi $ influences different physical observables so that BHs of two theories could be distinguished. One physical observable is the deflection angle of light in the KSBH and the influence of $\xi$ on this has recently been worked out by Uniyal, Nandan and Jetzer \cite{uniyal:2018}. An interesting recent work by Li and Deng \cite{li:2017} considered  photons coupling to Weyl tensor characterized by a parameter $\alpha$ and studied its influence in the Solar System for the deflection of light, the gravitational time delay and the Cassini experiment. Another excellent physical observable is the new diagnostic of Relative Time Delay (RTD)\footnote{We would like to submit that RTD is completely different from the well known Shapiro gravitational time delay. Interestingly, there is also an effect of \textit{gravitational\ time advancement}, proposed originally in \cite{bhadra:2010}, that could pass for an independent test of gravity theories. For an interesting development on this effect under gravity's rainbow, see \cite{xue:2017}.} of signals, denoted here by $\Delta t$, caused by the frame dragging of the spinning lens, which is the object of this paper. The stage to look for the RTD effect is set by the recent speculation of different binary lens systems in which we assume KSBH to be the spinning lens partner.

The concept of RTD as a potential diagnostic for gravity theories has so far been investigated rather scantily in the literature. The concept is as follows: Consider a binary system, where a variable light source $S$ orbits a spinning compact object (BH lens) \cite{dymnikova:1986,laguna:1997}: Suppose two light rays emanate from $S$ behind a spinning lens $L$ (with mass $M$, spin $a$), pass on either side of it to reach the observer at $O$, say, the Earth. The rays will reach at different times at $O$ and this difference in the times of arrival (TOA) is caused by the frame dragging effect due to the intervening spinning lens. The frame dragging causes light path lengths on either side of the lens to differ, shorter on the co-rotating side and longer on the counter-rotating side (see Fig.1). The RTD\ could be somewhat regarded as an astrophysical analogue of the quantum Bohm-Aharonov effect although the analogy is not too accurate since the light rays anyway pass through the weak gravitational field\footnote{For a true gravitational Bohm-Aharonov effect, see \cite{ford:1981}.}. The TOA of pulses has been calculated to the zeroth order by Laguna and Wolszczan in the Kerr metric for some hypothetical binary systems \cite{laguna:1997}. An effect of similar, though not exactly the same, nature was studied by Datta and Kapoor \cite{datta:1985}, where light was assumed to emerge not from a variable source behind the lens but from two diametrically opposite points on the spinning compact object itself that reached the observer on Earth. A potential example of RTD could be the early observation of extremely rapid fluctuations in the brightness of quasar $1525+227$ with characteristic time scale $\sim 200$ sec speculated to be caused by a spinning black hole of mass $M\sim 5\times 10^{8}M_{\odot }$ situated between the quasar and the observer \cite{matilsky:1982}. Recently, RTD has been studied for the Johannsen metric \cite{izmailov:2019} as a possible observable diagnostic to test the validity of the so-called "no-hair" conjecture of Penrose \cite{penrose:0}, which broadly states that only mass, electric charge and spin survive the collapse of baryonic matter to a black hole with the other degrees of freedom radiated away. For a recent review of the status of the conjecture, see Cuzinatto \textit{et al.} \cite{cuzinatto:2018}. So far no concrete experimental data on RTD are available but very accurate data can in principle constrain the values of $\xi $ that will have implications for no-hair conjecture as well as for the string theory.

The next concept to be employed is the thin-lens approximation (see the details, e.g., in Hartle \cite{hartle:2003}), where source, lens and observer are all considered as points and the light deflection takes place only at the lens while elsewhere the rays are assumed to travel in straight lines. This is an excellent approximation when the rays travel vast distances compared to the size of the lens with the impact parameter far larger than the lens size. Therefore, the relevant angles with the optical axis $SLO$ are small, making the quadrilateral in Fig.1 slender. We note that a more accurate calculation of RTD should consider exact null geodesics around the spinning lens and their numerical integration should be performed between two finite points on the optical axis. Nonetheless, the thin-lens and the ensuing PPN approximation provides a simple and elegant descripion of many realistic lensing situations \cite{hartle:2003}. might not mean much of a compromise with rigor while, on the contrary, would suffice to capture the leading order values of the RTD observable, as we will soon see, in the same manner as the usual PPN observables in the Schwarzschild spacetime.

The purpose of this paper is to theoretically study the effect of $\xi$ on RTD in the binary lens systems assuming an alignment of source, lens and observer. Specifically, we want to see how $\Delta t$ differs in magnitude from the Kerr value ($\xi =0$) for a given lens, when its mass $M$ and spin $a$ are independently specified. We wish to generalize the components of $\Delta t$, including the known zeroth order law, to finite distance lens scales that are practically realizable. We shall argue that the spinning lens can be considered equivalent to an "effective" static lens\footnote{This result is similar to the result of a "displaced Schwarzschild lens" equivalent to a Kerr lens up to first order in spin, as shown generically by Sereno \cite{sereno:2002}.}, which would offer us an opportunity to measure the zeroth order RTD in terms of the observable image magnification factor. Numerical values will be tabulated for two speculated binary lens systems. The method adopted in this paper can be applied with ease to other spinning metrics available in the literature (see, e.g., \cite{manko:1992,glampedakis:2006,bambi:2013,ghosh:2015,toshmatov:2014,toshmatov:2017}). We shall take $G=1$, $c=1$ unless specifically restored.

The paper is organized as follows. In Sec.2, we shall rewrite the KSBH for our purpose and in Sec.3, derive the equations for the RTD. In Sec.4, we shall integrate the straight null path equations. Sec.5 contains the idea of the equivalent static lens system and the corresponding observable magnification of images. Sec.6 presents indicative numerical estimates for two binary lens systems and Sec.7 concludes the paper.

\section{Kerr-Sen black hole (KSBH)}
\label{sec:ksbh}
The string-theory 4d effective action is \cite{sen:1992}
\begin{eqnarray}
S &=&-\int d^{4}x\sqrt{-\mathcal{G}}e^{-\Phi}\times\left(-\mathcal{R+}\frac{1}{12}\mathcal{H}_{\mu\nu\rho}\mathcal{H}^{\mu\nu\rho}\right. \nonumber \\
&&\left. -\mathcal{G}^{\mu \nu}\partial_{\mu}\Phi \partial_{\nu}\Phi +\frac{1}{8}\mathcal{F}_{\mu\nu}\mathcal{F}^{\mu\nu}\right),
\end{eqnarray}%
where $\mathcal{G}_{\mu\nu}$ is the metric in the string frame, $\mathcal{R}$ is the Ricci scalar, $\mathcal{F}_{\mu\nu} = \partial_{\mu}\mathcal{A}_{\nu} - \partial_{\nu}\mathcal{A}_{\mu}$ is the field strength corresponding to the Maxwell field $\mathcal{A}_{\mu}$, $\Phi $ is the dilaton field and
\begin{eqnarray}
\mathcal{H}_{\mu\nu\rho} &=&\partial_{\mu}\mathcal{B}_{\nu\rho} + \partial_{\nu}\mathcal{B}_{\rho\mu} + \partial_{\rho}\mathcal{B}_{\mu\nu} \nonumber \\
&&- \frac{1}{4}\left(\mathcal{A}_{\mu}\mathcal{F}_{\nu\rho} + \mathcal{A}_{\nu}\mathcal{F}_{\rho\mu} + \mathcal{A}_{\rho}\mathcal{F}_{\mu\nu}\right),
\end{eqnarray}%
where the last term is the gauge Chern-Simons term and $\mathcal{B}_{\mu\nu}$ is an antisymmetric tensor gauge field.

The KSBH solution \cite{sen:1992} in Boyer-Lindquist coordinates ($t$, $r$, $\theta$, $\phi$) in the Einstein frame, with the redefined metric $g_{\mu\nu}=e^{-\Phi}\mathcal{G}_{\mu\nu}$, is given by
\begin{equation}
d\tau^{2} = g_{tt}dt^{2} + g_{rr}dr^{2} + g_{\theta\theta}d\theta^{2} + g_{\phi\phi}d\phi^{2} + 2g_{t\phi}dtd\phi,
\end{equation}
\begin{eqnarray}
g_{tt} &=& -\left(\frac{\Delta - a^{2}\sin^{2}{\theta}}{\Sigma}\right), \\
g_{rr} &=& \frac{\Sigma}{\Delta}, \\
g_{\theta\theta} &=& \Sigma, \\
g_{\phi\phi} &=& \left(\frac{\Xi}{\Sigma}\right)\sin^{2}{\theta }, \\
g_{t\phi} &=& -\frac{2mar\cosh^{2}{\alpha} \sin^{2}{\theta}}{\Sigma}, \\
\mathcal{A}_{t} &=& \frac{mr\sinh{2\alpha}}{\sqrt{2}\Sigma}, \; \mathcal{A}_{\phi}=\frac{mar\sinh{2\alpha}\sin^{2}{\theta}}{\sqrt{2}\Sigma} \\
\mathcal{B}_{t\phi} &=& \frac{2a^{2}mr\sinh^{2}{\alpha}\sin^{2}{\theta}}{\Sigma}, \\
\Phi &=& -\frac{1}{2}\ln{\left(\frac{\Sigma}{r^{2}+a^{2}\cos^{2}{\theta }}\right)} ,
\end{eqnarray}%
where the symbols are
\begin{eqnarray}
\Delta &=& r^{2} + a^{2} - 2mr, \\
\Sigma &=& r^{2} + a^{2}\cos^{2}{\theta} + 2mr\sinh^{2}{\alpha}, \\
\Xi &=& \left(r^{2} + a^{2} + 2mr\sinh^{2}{\alpha}\right)^{2} - a^{2}\Delta\sin^{2}{\theta}.
\end{eqnarray}%
The parameters $m$, $\alpha$ and $a$ are related to the black hole mass $M$, dilatonic charge $Q$ and the angular momentum $J$ as follows \cite{uniyal:2018}
\begin{equation*}
M = \frac{m}{2}(1+\cosh{2\alpha}),\; Q = \frac{m}{\sqrt{2}}\sinh{2\alpha},\; J = \frac{am}{2}(1+\cosh{2\alpha}).
\end{equation*}%
Here, for our convenience, by defining the string parameter $\xi$ as
\begin{eqnarray}
\xi &\equiv& m\sinh^{2}{\alpha}, \\
m &=& M-\xi .
\end{eqnarray}%
we rewrite the metric in the form
\begin{eqnarray}
g_{tt} &=& -\left(1-\frac{2Mr}{\Sigma}\right), \\
g_{rr} &=& \frac{\Sigma}{\Delta}, \\
g_{\theta\theta} &=& \Sigma, \\
g_{\phi\phi} &=& \left[r(r+2\xi) + a^{2} + \frac{2Mra^{2}\sin^{2}{\theta}}{\Sigma}\right]\sin^{2}{\theta}, \\
g_{t\phi} &=& -\frac{2Mar\sin^{2}{\theta}}{\Sigma},
\end{eqnarray}%
where
\begin{eqnarray}
\Sigma &=& r(r+2\xi) + a^{2}\cos^{2}{\theta}, \\
\Delta &=& r(r+2\xi) + a^{2} - 2Mr, \\
\xi &=& \frac{Q^{2}}{2M}.
\end{eqnarray}

The outer and inner event horizons appear at ($\Delta =0$):%
\begin{equation}
r_{\pm}=\left(M-\xi\right) \pm \sqrt{\left(M-\xi \right)^{2}-a^{2}},
\end{equation}%
which are real if $a\leq\left(M-\xi\right)$, providing an upper limit on $\xi$ as
\begin{equation}
\xi \leq M(1-a_{\ast }),
\end{equation}%
where $a_{\ast}\left(\equiv\frac{a}{M}\right)$ is the non-dimensionalized spin parameter. Independent measurements of $M$ and $a_{\ast}$ for a given spinning black hole automatically provides an upper limit on $\xi$. The Page-Thorne maximum value for a spinning black hole is $a_{\ast}=0.998$, which implies
\begin{equation}
\xi \leq 0.002M.
\end{equation}%
The equality in the above implies the extremal value of $\xi$. Only very accurate pico-second level data can constrain $\xi$ more strigently, not excluding even the value $\xi =0$. One such possible experiment could be the measurement of RTD described below.

\section{Relative Time Delay (RTD)}
\label{sec:rtd}
To derive the equation for RTD, consider the null trajectory on the equatorial plane ($\theta=\pi/2$)\footnote{%
Light motion on the equatorial plane, which is responsible for RTD,
is possible in the Kerr--Sen black hole. In fact, the bending of light on
the equatorial plane ($\theta =\pi /2$) of the Kerr-Sen BH has recently been
calculated in \cite{he:2016}.} given by $d\tau^{2}=0$, so that the coordinate time required for light rays along an infinitesimal null world line is given by
\begin{equation}
dt_{\pm} = \frac{d\phi}{g_{tt}}[-g_{t\phi }\pm h(r,\phi )],
\end{equation}%
where
\begin{equation}
h(r,\phi )\equiv \sqrt{g_{t\phi }^{2}-g_{tt}\left\{g_{rr}\left(\frac{dr}{d\phi}\right)^{2} + g_{\phi\phi}^{2}\right\}}.
\end{equation}%
\begin{figure}[!ht]
  \includegraphics[width=0.50\textwidth]{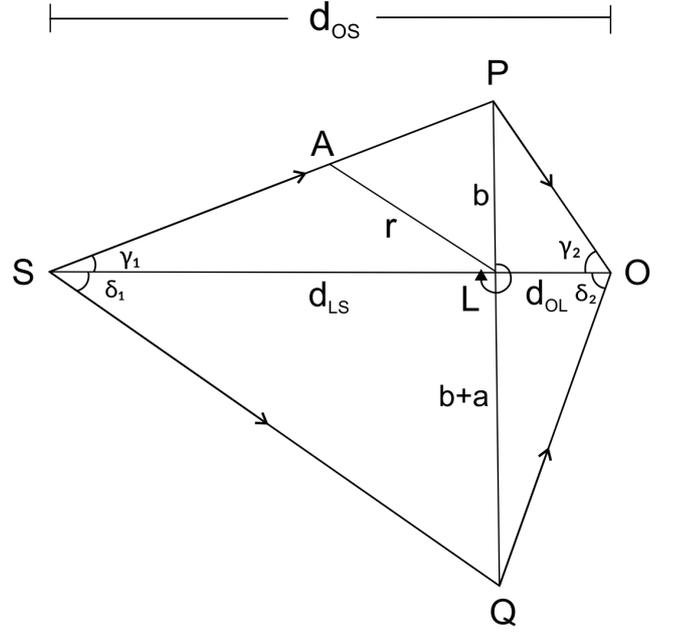}
\caption{The generic thin-lens slender quadrilateral (here angles are exaggerated). $S$, $L$ and $O$ are the source, lens and observer respectively aligned on a straight line ($\beta =0$), $b$ is the impact parameter and $a$ is the spin. The arbitrary angles are as shown.}
\label{fig:1}
\end{figure}
We assume the passage of coordinate time to be positive for both $\pm$ sides of the lens identifying $d\phi >0$, for light rays passing the lens by the co-rotating side ($+$) and $d\phi<0$ for the counter-rotating side ($-$), so that $dt_{+}$ and $dt_{-}$ are both positive. The net difference between the two null rays in the time of arrival (TOA) at the observer is also positive and is given by:
\begin{eqnarray}
dt=dt_{-}-dt_{+}&=&\frac{\left\vert d\phi\right\vert}{g_{tt}}[g_{t\phi}+h(r,\phi)] - \frac{\left\vert d\phi\right\vert}{g_{tt}}[-g_{t\phi}+h(r,\phi)] \nonumber\\
&=& \frac{2g_{t\phi}}{g_{tt}}\left\vert d\phi\right\vert.
\end{eqnarray}%
This delay $dt$ is due to the frame-dragging effect characterized by$\left(\frac{2g_{t\phi}}{g_{tt}}\right)$, which we are going to compute in this paper. We assume that the source, spinning lens and the observer are aligned, that is, they are situated on a straight line (see Fig.1). When the lens is not spinning, the path lengths of the light ray on both sides of the lens would be the same and there would be no difference in TOA at the observer. However, when the lens is spinning the path lengths will differ - longer for co-rotating and shorter for counter-rotating rays - giving rise to the RTD. The rays on the equatorial plane pass are required to pass through the weak field so that the thin-lens approximation is valid, that is the distance of closest approach $r>>r_{ph}^{\pm}$ on either side of the lens. This large $r$ ensures that, for a given lens of mass $M$ and spin $a$, the quantitiy $\left(\frac{M}{r}\right)<<1$.

We shall apply the generic formula (30) to KSBH by expanding $\left(\frac{2g_{t\phi}}{g_{tt}}\right)$ such that, to first order in $a$ and up to third PPN order in $\left( M/r\right)$, the expansion in the curly bracket below is justified:
\begin{equation}
dt=|d\phi |\left( \frac{1}{c}\right) \left[ \frac{4aM}{r}\left\{ 1+\frac{2M}{r}\lambda _{1}+\frac{M^{2}}{2r^{2}}\lambda _{2}+...\right\} \right],
\end{equation}
where%
\begin{eqnarray}
\lambda _{1}&=&1-\frac{Q^{2}}{2M^{2}}=1-\frac{\xi}{M},  \\
\lambda _{2}&=&8-\frac{8Q^{2}}{M^{2}}+\frac{2Q^{4}}{M^{4}} \nonumber\\
&=& 8\left( 1-\frac{2\xi }{M}+\frac{\xi^{2}}{M^{2}}\right) =8\left( 1-\frac{\xi}{M}\right)^{2}
\end{eqnarray}
are the parameters showing deviations away from Kerr BH having values $\lambda_{1}=1$, $\lambda_{2}=8$. Note that the factor $\left(\frac{4aM}{r}\right)$ multiplying the curly bracket in Eq.(31) need not be small. The effect of $\xi $ is evident from above and for a given mass $M$ and spin $a$, the measured data on $dt$ would in principle constrain $\xi$.

The total RTD $\Delta t$ between two null rays traveling from the source to observer along two opposite sides of the intermediate spinning lens is%
\begin{eqnarray}
\Delta t &=&\left( \frac{1}{c}\right) \int\limits_{0}^{\pi }d\phi \left[
\frac{4aM}{r}\left\{ 1+\frac{2M}{r}\lambda _{1}+\frac{M^{2}}{2r^{2}}\lambda
_{2}+...\right\} \right] \\
&\equiv &\frac{1}{c}\left( I_{1}+I_{2}+I_{3}\right) =\Delta t_{1}+\Delta
t_{2}+\Delta t_{3}.
\end{eqnarray}%
We compute the integral locating the spinning lens at the origin of a polar system of coordinates on the equatorial plane ($\theta =\pi/2$). As can be seen, $\xi$ does not influence the leading first order RTD $\Delta t_{1}$. In the following, we shall derive explicit expressions for $\Delta t_{1}$, $\Delta t_{2}$ and $\Delta t_{3}$ within the thin-lens approximation.

\begin{figure}[!ht]
  \includegraphics[width=0.50\textwidth]{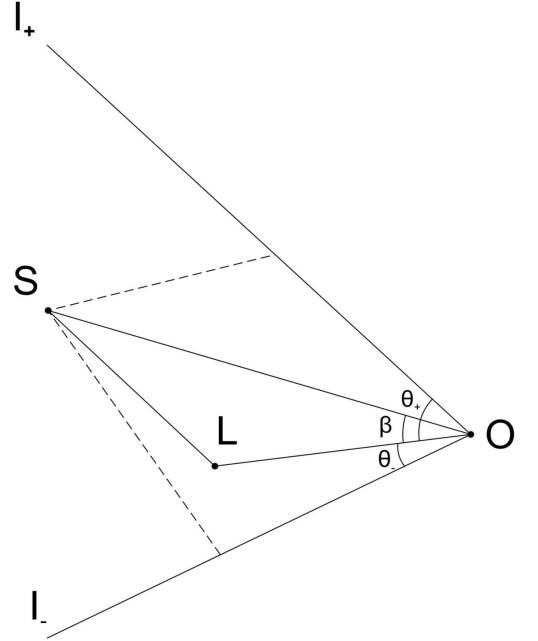}
\caption{"Effective" non-aligned static lens with $\beta\neq 0$ equivalent to the aligned spinning lens of Fig.1 with $\beta=0$. The images appear at $I_{\pm}$ at two angular locations $\theta_{\pm}$.}
\label{fig:2}
\end{figure}

\section{Thin-lens approximation}
\label{sec:tla}
In realistic lensing configurations, the radius over which bending takes place is of the order of Schwarzschild radius, which is much smaller than the typical distances $d_{OL}$, $d_{OS}$, $d_{LS}$ over which light propagates. Then, to an excellent approximation, light rays can be assumed to propagate as straight lines when far away from the lens, with the bending taking place only at the point lens \cite{hartle:2003}. In a spinning lens situation, we argue that the approximation requires three additional non-trivial conditions to be satisfied as enumerated below.

(i) The first condition is that the rays emerging from $S$, after passing along line segments on either side of spinning lens $L$ in the equatorial plane, should meet exactly at $O$ making a quadrilateral $SPOQ$ of Fig.1. We show that this is possible only under the assumption that $(a/b)<<1$ such that orders of $(a/b)^{2}$ and higher can be neglected. From Fig.1, the angles are related by%
\begin{equation}
\gamma _{1}+\gamma _{2}+\measuredangle OPS =\pi =\measuredangle OPS+\theta
_{1}\Rightarrow \gamma _{1}+\gamma _{2}=\theta _{1},
\end{equation}%
\begin{equation}
\delta _{1}+\delta _{2}+\measuredangle OPS =\pi =\measuredangle OPS+\theta
_{2}\Rightarrow \delta _{1}+\delta _{2}=\theta _{2}.
\end{equation}%
In the thin-lens approximation, the relevant angles given below
\begin{equation}
\gamma_{1}=\frac{b}{d_{LS}},\; \gamma_{2}=\frac{b}{d_{OL}},\; \delta_{1}=\frac{b+x}{d_{LS}},\; \delta_{2}=\frac{b+x}{d_{OL}},
\end{equation}
should be small, where $x(a,b)$ is an unknown function to be determined. For a given $d_{OL}$, the impact parameter $b$ and the ratio $\chi$ ($=\frac{d_{LS}}{d_{OL}}$) be such that the angles $\gamma_{1}$, $\gamma_{2}$, $\delta_{1}$, $\delta_{2} <<1$.

Following the Boyer-Lindquist formula \cite{boyer:1967} for the bending of light at $P$ and $Q$ respectively, we have%
\begin{eqnarray}
\theta _{1} &=&\frac{4M}{b}\left( 1-\frac{a}{b}\right) ,\text{ (prograde motion of light)} \\
\theta _{2} &=&\frac{4M}{b}\left( 1+\frac{a}{b+x}\right) ,\text{ (retrograde motion of light)}
\end{eqnarray}%
Also%
\begin{eqnarray}
b\left( \frac{1}{d_{LS}}+\frac{1}{d_{OL}}\right) &=&\theta _{1}, \\
\left( b+x\right) \left( \frac{1}{d_{LS}}+\frac{1}{d_{OL}}\right) &=&\theta
_{2.}
\end{eqnarray}%
Hence%
\begin{equation}
\frac{b}{b+x}=\frac{1-\frac{a}{b}}{1+\frac{a}{b+x}}=\frac{\theta _{1}}{%
\theta _{2}}.
\end{equation}%
The first equality yields%
\begin{equation}
\frac{(x+2b)\left[ -bx(x+b)+a(b^{2}+bx+x^{2})\right] }{b^{2}(b+x)(a+b+x)}=0.
\end{equation}%
This equation is valid when $x\neq -b$ and $-(a+b)$. We see that $x=-2b$ satisfies the above equation but it gives the closest approach distance $b+x\rightarrow -b$, which is ruled out since distance cannot be negative. Therefore, we are left with the quadratic equation in $x(a,b)$ as
\begin{equation}
-bx(x+b)+a(b^{2}+bx+x^{2})=0.
\end{equation}%
Solving the equation under the condition that $x(a,b)=0$ at $a=0$, we find that the only root that satisfies this condition is%
\begin{eqnarray}
x_{1} &=&\frac{(-ab+b^{2})-\sqrt{b^{4}+2ab^{3}-3a^{2}b^{2}}}{2(a-b)}  \notag \\
&=&a\left[ 1+\left( \frac{a}{b}\right) ^{2}-\left( \frac{a}{b}\right)^{3}+...\right].
\end{eqnarray}%
Neglecting $\left(\frac{a}{b}\right)^{2}$ and higher orders, we get $x_{1}=a$, yielding the impact parameters $b$ and $b+a$ as in Fig.1.

(ii) The second condition is that the light rays should pass far away from the spinning lens that their trajectories can be approximated by straight lines. To determine how far is far, we need to know the critical values of the closest approach distance $r_{\text{cr}}^{\pm}$ or the radii of the photon spheres $r_{\text{ph}}^{\pm}$ appearing respectively on the co-rotating ($+$) and counter-rotating sides ($-$) of the lens. For the KSBH, they have been derived by Uniyal, Nandan and Jetzer \cite{uniyal:2018}:%
\begin{equation}
r_{\text{ph}}^{\pm }=\xi +\frac{2}{3}\left( 3M-\xi \right) \left[ 1+\cos
\left\{ \frac{2}{3}\arccos \left( \frac{\mp 3a}{3M-\xi }\sqrt{\frac{3M}{%
3M-\xi }}\right) \right\} \right] .
\end{equation}%
For the Kerr BH, $\xi =0$, and $r_{\text{ph}}^{\pm}$ exactly coincide with the radii of the photon spheres derived by \cite{bardeen:1973}:
\begin{equation}
r_{\text{ph}}^{\pm }=2M\left[ 1+\cos \left\{ \frac{2}{3}\arccos \left( \frac{%
\mp a}{M}\right) \right\} \right] .
\end{equation}%
When $a=0$, $r_{\text{ph}}^{\pm\text{KSBH}}=3M$ even though the KSBH metric does not reduce to Schwarzschild but to Gibbons-Maeda-Garfinkle-Horowitz-Strominger (GMGHS) dilatonic static BH \cite{gibbons:1988,garfinkle:1991}, although $r_{\text{ph}}^{\pm\text{KerrBH}}=r_{\text{ph}}^{\text{SchBH}}=3M$. We recall that at the photon spheres around BHs, the deflection angles become logarithmically divergent, as a result of which the light rays get eternally captured there \cite{manko:1992,glampedakis:2006}. Consequently, these radii $r_{\text{ph}}^{\pm}$ demarcate natural strong field limits perceived by null rays. Therefore, light rays should pass at distances far away from the radii $r_{\text{ph}}^{\pm \text{KSBH}}$ to ensure that the angles (in radian) be small compared to unity.

(iii) Thin lens approximation breaks down when $\frac{M}{b}\sim O\left(1\right)$. Therefore, to avoid it, the third condition is that the smaller of the two impact parameters, $b$ and $b+a$, must far exceed the larger of the two radii $r_{\text{ph}}^{\pm}$ of photon spheres, viz., $b>>r_{\text{ph}}^{-}$ or let $b=10^{n}r_{\text{ph}}^{-}$, where $n>1$ is any real number. Our idea is to march $r$ towards $r_{\text{ph}}^{-}$ to the extent that the rays preserve the smallness of the involved angles in Fig.1, as discussed in (i). This algorithm will be exercised in Tables 1 and 2 below.

There is a fourth mandatory condition for KSBH that the value of $\xi/M$ be actually very tiny respecting the extreme limit in (26). Accordingly, for illustration, we shall assume $0\leq\xi/M\leq 10^{-3}$ (say). Returning to Fig.1, we have by construction $d_{LS}=\chi d_{OL}$, where $\chi>0$ is a finite constant, $PLQ\perp OLS$, and the arbitrary angles are as indicated. The radial distance is measured from the lens $L$. By piecewise integration of the straight lines in the relevant sectors, corotating $\left(\frac{1}{r_{\text{cor}}}\right)$ and counterrotating $\left(\frac{1}{r_{\text{cou}}}\right)$ sides, we get the final result by subtracting the integrals between the path lengths, viz., $SQO-SPO$:%
\begin{eqnarray}
I_{1} &=& 4aM\left[\int\limits_{\pi/2}^{\pi}\frac{1}{r_{\text{cor}}}d\phi +\int\limits_{0}^{\pi/2}\frac{1}{r_{\text{cor}}}d\phi\right] \nonumber\\
&&- 4aM\left[ \int\limits_{-\pi /2}^{-\pi }\frac{1}{r_{\text{cou}}}d\phi +\int\limits_{0}^{-\pi /2}\frac{1}{r_{\text{cou}}}d\phi \right],
\end{eqnarray}%
\begin{eqnarray}
\Delta t_{1} =\frac{I_{1}}{c} &=& \frac{8a^{2}M\left\{b\left(\chi - 1\right) +\chi d_{OL}\right\}}{cb(a+b)\chi d_{OL}} \nonumber\\
&&+ \frac{8aM\left\{ b\left( \chi -1\right) +2\chi d_{OL}\right\}}{c(a+b)\chi d_{OL}}.
\end{eqnarray}%
By taking the limit where the source and observer are both infinite distance away from the lens, that is when $d_{OL}\rightarrow\infty$, which also implies $d_{LS}\rightarrow\infty$, we have
\begin{equation}
\Delta t_{1}=\frac{8aM(a+2b)}{cb(a+b)}.
\end{equation}
The requirement of thin-lens approximation further implies that $(a/b)<<1$ such that orders of $(a/b)^{2}$ and higher can be neglected, then we end up with
\begin{equation}
\Delta t_{1}\simeq \frac{16aM}{cb},
\end{equation}
which is independent of $\xi$ and is precisely the leading order delay obtained by Laguna \& Wolszczan \cite{laguna:1997}. Interestingly, Eq.(53) exactly follows from Eq.(51) when $\chi =1$, that is, when $d_{LS}=d_{OL}$. Thus Eq.(51) generalizes the delay (53) to realistic \textit{finite} distance thin-lens geometry. In the same way, we can calculate the integrals $I_{2}$ and $I_{3}$ which, to leading orders in $\left( \frac{M}{b}\right) ^{2}$ and $\left(\frac{M}{b}\right)^{3}$, work out to
\begin{eqnarray}
\Delta t_{2} &=&\frac{I_{2}}{c}=\frac{4aM^{2}\left[a^{2}F_{1}+2abF_{2}+b^{2}F_{3}\right] }{cb^{2}(a+b)^{2}\left(\chi d_{OL}\right)^{2}}\lambda_{1}, \\
\Delta t_{3} &=&\frac{I_{3}}{c}=\frac{2aM^{3}\left[ G_{1}+G_{2}-G_{3}-G_{4}\right] }{3cb^{3}d_{OL}^{3}}\lambda _{2}
\end{eqnarray}%
where%
\begin{eqnarray}
F_{1} &\equiv &2b(\chi -1)\chi d_{OL}+\pi \left(\chi d_{OL}\right)^{2}+\pi b^{2}\left( 1+\chi ^{2}\right) , \\
F_{2} &\equiv &3b(\chi -1)\chi d_{OL}+\pi \left(\chi d_{OL}\right)^{2}+\pi b^{2}\left( 1+\chi ^{2}\right) , \\
F_{3} &\equiv &4b(\chi -1)\chi d_{OL}+2\pi \left(\chi d_{OL}\right)^{2}+\pi b^{2}\left( 1+\chi ^{2}\right) , \\
G_{1} &\equiv &\left[ b+d_{OL}\right] \left(2b^{2}+bd_{OL}+2d_{OL}^{2}\right) , \\
G_{2} &\equiv &\left[ \frac{a+b+d_{OL}}{\left( 1+a/b\right) ^{3}}\right]\left[ 2\left( a+b\right) ^{2}+\left( a+b\right)d_{OL}\right. \nonumber\\
&&\left.+2d_{OL}^{2}\right],
\end{eqnarray}
\begin{eqnarray}
G_{3} &\equiv &\left[ \frac{b-\chi d_{OL}}{\chi ^{3}}\right] \left[2b^{2}-b\chi d_{OL}+2(\chi d_{OL})^{2}\right] , \\
G_{4} &\equiv &\left[ \frac{1}{\left( 1+a/b\right) ^{3}\chi ^{3}}\right]\left[ 2\left( a+b\right) ^{3}-3\left( a+b\right)^{2}\chi d_{OL}\right. \nonumber \\
&&\left.+ 3\left(a+b\right)\left(\chi d_{OL}\right)^{2}-2(\chi d_{OL})^{3}\right].
\end{eqnarray}

The Eqs. (50,53,54) are the generic equations for the delay. We shall plug the lens parameter values $a,M$, distance values $b$, $d_{OL}$, $\chi=d_{LS}/d_{OL}$ into these equations, take care to preserve the smallness of the angles $b/d_{OL}$, $b/d_{LS}<<1$, and tabulate the RTD components $\Delta t_{1}$, $\Delta t_{2}$ and $\Delta t_{3}$, the last two containing the the string parameter $\xi$ via $\lambda_{1}$ and $\lambda_{2}$ (Tables 1,2).

\section{Image magnification and RTD}
\label{sec:imrtd}
We briefly argue here that the RTD can be experimentally measured in terms of the observable image magnification determined by $\beta/\theta_{E}$, where $\theta_{E}$ is the Einstein angle. The idea is to imagine the aligned spinning lens system to be replaced by an "effective" static non-aligned lens system with $\beta\neq 0$. If the lens were truly static and aligned, there would be no frame dragging, the rays from the source would traverse equal lengths to the observer from the Einstein ring, so there would be no time delay. However, due to the spinning, and consequent frame dragging, the incoming rays from the source along the two different angles $\gamma_{2}$ and $\delta_{2}$ would traverse two different distances with a path length difference $\Delta D$, say. Instead of the Einstein ring, the rays passing by the effective static lens will now produce two distinct images $\theta_{\pm}$ situated asymmetrically on either side of the optical axis $OL$ and light from them would reach the observer with a time gap $\Delta D/c$. We need to calculate the ratio $\left(\frac{a}{b}\right)$ from observational data, assuming that the impact parameter $b=L/E$ is known.

The angular positions $\theta_{\pm}$ of the images follow from the effective static lens geometry in Fig.2:
\begin{equation}
\theta_{\pm} = \frac{1}{2}\left[\beta\pm\left(\sqrt{\beta^{2}+4\theta_{E}^{2}}\right)\right],
\end{equation}%
where $\theta_{E}$ is the Einstein angle defined by
\begin{equation}
\theta_{E}=\left(\frac{4Md_{LS}}{d_{OL}d_{OS}}\right)^{1/2}.
\end{equation}
The difference $\Delta D$ in path length in the static case, assuming $\beta << \theta_{E} << 1$, is \cite{hartle:2003}
\begin{equation}
\Delta D\simeq 2\beta \theta _{E}d_{OS}.
\end{equation}%
Under the simplifying assumption $d_{OL}=d_{LS}=\frac{d_{OS}}{2}$ (or, in our notation, $\chi =1$), the above equation reduces to
\begin{equation}
\Delta D/c\simeq \frac{1}{2}\left(\frac{\beta}{\theta_{E}}\right)\left(\frac{16M}{c}\right) \simeq \left(\frac{a}{b}\right) \left( \frac{16M}{c}\right) = \Delta t_{1}.
\end{equation}%
Comparing with $\Delta D/c$ the value for $\Delta t_{1}$ from Eq.(52) valid for $\frac{a}{b}<<1$, we get
\begin{equation}
\frac{a}{b} = \frac{1}{2}\left( \frac{\beta}{\theta_{E}}\right) <<1.
\end{equation}%
The approximation shown in Eq.(65) is due to thin-lens approximation that can be improved by considering the exact null ray path equation. Further details of lensing by Schwarzschild black hole can be found in Virbhadra \& Ellis \cite{virbhadra:2000}.

Note that the values of $\beta$ and $\theta_{E}$ determine the shapes and magnifications (or ratios of brightness) of the images. While the azimuthal width $\Delta\phi$ is unchanged, the polar width $\Delta\theta_{\pm}$ of the images change and its magnitude can be obtained by differentiating Eq.(62):
\begin{equation}
\Delta\theta_{\pm}=\left(\frac{1}{2}\right)\left[1\pm\frac{\beta}{\sqrt{\beta^{2}+4\theta_{E}^{2}}}\right]\Delta\beta .
\end{equation}
Since the angular width of the source $\Delta\beta\neq 0$, this result implies a distorted and elongated shape of the images that have been confirmed by observations.

The ratio of the brighness of the individual images $\mu_{\pm}$ to the unlensed brightness $\mu_{\ast}$ at the angular positions $\theta_{\pm}$ is given by the individual magnifications \cite{hartle:2003}
\begin{eqnarray}
\frac{\mu_{\pm}}{\mu_{\ast}} &=& \left\vert\left(\frac{\theta_{\pm}}{\beta}\right)\left(\frac{d\theta_{\pm}}{d\beta}\right)\right\vert \notag \\
&=&\frac{1}{4}\left[\frac{\beta}{\sqrt{\beta ^{2}+4\theta_{E}^{2}}}+\frac{\sqrt{\beta^{2}+4\theta_{E}^{2}}}{\beta}\pm 2\right].
\end{eqnarray}
We can draw a very interesting conclusion from here: Since $x+1/x\geq 2$, we can conclude that, for $\beta <0$, the brightness $\left\vert\mu_{-}\right\vert > \mu_{+}$ showing that the image $\theta_{-}$ on the counter-rotating side is brighter than the image $\theta_{+}$ on the co-rotating side. This observation alone would indicate that the intervening lens is spinning! However, the magnitudes of individual image brightness $\mu_{\pm}$ for any given source do not differ very greatly thus making the individual measurements difficult. In this case, another very useful quantity is the total magnification over the background $\mu_{\ast}$, called the \textit{magnification factor }defined by \cite{hartle:2003}
\begin{eqnarray}
\frac{\mu_{\text{tot}}}{\mu_{\ast}} &=& \frac{\mu_{+}-\mu_{-}}{\mu_{\ast}} = \frac{1}{2}\left\vert\frac{\beta}{\sqrt{\beta^{2} + 4\theta_{E}^{2}}} + \frac{\sqrt{\beta^{2} + 4\theta_{E}^{2}}}{\beta}\right\vert \notag \\
&=&\frac{1}{2}\left\vert\frac{\left(\beta/\theta_{E}\right)}{\sqrt{\left(\beta/\theta_{E}\right)^{2}+4}}+\frac{\sqrt{\left(\beta /\theta_{E}\right)^{2}+4}}{\left(\beta/\theta_{E}\right)}\right\vert,
\end{eqnarray}%
the ratio being always greater than unity. Especially, when the effective $\left(\frac{\beta}{\theta_{E}}\right) $ is small, that is, the source is very close to the optical axis $OL$, the total magnification could be quite large that should be observable. The measurement of magnification can lead to the determination of $\left(\frac{\beta}{\theta_{E}}\right)$ and hence of $\Delta t_{1}$. Below we tabulate some numerical estimates of RTD for typical lensing binaries speculated in the literature.

\section{Numerical estimates}
\label{sec:num}
Pulsar-BH (PSR-BH) binary systems provide the best laboratory for testing the RTD predictions since variable sources like pulsars, quasars, GRBs etc can give out signals at the instant they are behind the spinning black hole on the optical axis $OL$ and their times of arrival $\Delta t$ can be measured at the observer. Though a concrete example of such a binary is yet to be detected, the prospects for discovery seem quite promising \cite{laguna:1997}. We assume the pulsar orbit to be in the equatorial plane of the KSBH and the line of sight is perpendicular to the axis of the KSBH spin.

\begin{center}
\textit{(a) PSR-Cygnus X-1 binary}
\end{center}

An early estimate was that, of all pulsars discovered so far, a small but significant number of them belong to a PSR-BH category with a BH having masses a few times of solar masses \cite{lipunov:2005}. We consider a typical illustration, namely, of a PSR-Cygnus X-1 binary with $M=14.8M_{\odot}=2.19\times 10^{6}$ cm, $a=0.95M=2.08\times 10^{6}$ cm \cite{gou:2011}, $d_{OL}=1.86$ kpc $=5.74\times 10^{21}$ cm \cite{reid:2011}. The Kerr BH case corresponds to $\xi=0$, $\lambda_{1}=1$, $\lambda_{2}=8$. The value of $r_{\text{ph}}^{\pm}$ in Eq.(47) is practically insensitive to the small values of $\xi/M$ chosen respecting the upper limit given in (26), and one has $r_{\text{ph}}^{-}=8.66\times 10^{6}$ cm and $r_{\text{ph}}^{+}=3.03\times 10^{6}$ cm, so $r_{\text{ph}}^{-}$ is the larger of the two radii. Accordingly, to preserve the thin-lens and PPN approximation, we choose $b=10^{4}r_{\text{ph}}^{-}=8.66\times 10^{10}$ cm, so that $\frac{M}{b}\sim 10^{-5}$ justifying the PPN expansion. The lens-source distances $d_{LS}=\chi d_{OL}$ in Fig.1 are varied by varying $\chi$ but preserving the required smallness of the angles (in radian): $\gamma_{1}\simeq b/d_{LS}$, $\delta_{1}\simeq (a+b)/d_{LS}$, $\gamma_{2}\simeq b/d_{OL}$, $\delta_{2}\simeq (a+b)/d_{OL}$. Then, one has $\Delta t_{1}=2.8\times 10^{-8}$ sec $=0.028$ $\mu$sec. Thus, even the impact parameter $b$ is ten thousand times farther than $r_{\text{ph}}^{+}$, Table 1 shows that, to leading order, RTD component $\Delta t_{1}$ would be at the $\mu$sec level that could be measurable in future. The other components containing $\xi$ are $\Delta t_{2}$, $\Delta t_{3}$, which hold no promise to be measurable even in the far future.

\begin{table}
\centering
\begin{tabular}{|c|c|c|c|c|}
\hline
$\xi /M$ & $\chi $ & $\Delta t_{1}$ ($\mu $sec) & $\Delta t_{2}$ ($\mu $sec) & $\Delta t_{3}$ ($\mu $sec) \\
\hline
$0$ & $0.1$ & $0.028$ & $1.11\times 10^{-6}$ & $5.97\times 10^{-12}$  \\
$10^{-5}$ & $0.1$ & $0.028$ & $1.00\times 10^{-6}$ & $4.84\times 10^{-12}$ \\
$10^{-4}$ & $0.1$ & $0.028$ & $5.57\times 10^{-7}$ & $1.49\times 10^{-12}$ \\
\hline
\end{tabular}
\caption{The table shows some typical values of RTD components $\Delta t_{1}$, $\Delta t_{2}$, $\Delta t_{3}$ for the PSR-Cygnus X-1 binary. The last two columns contain the effect of the string parameter $\xi$ ($=Q^{2}/2M$) through $\lambda_{1}$ and $\lambda_{2}$ as in Eqs.(32) and (33). The distances $d_{LS}$, $d_{OL}$ in Fig.1 are such that the angles remain small: $\gamma_{1} \simeq \tan{\gamma_{1}} = b/d_{LS} \simeq \delta_{1} \simeq \tan{\delta_{1}}$ etc. The Kerr values are at $\xi =0$ (first row). The values of $\xi/M$ are chosen respecting the upper limit (26).}
\end{table}

High precision measurement is best possible in the case of millisecond pulsars, and a precision of $0.1$ $\mu$sec was achieved by \cite{vanStraten:2001} for $PSRJ0437-4715$, a bright millisecond pulsar in a White Dwarf-Neutron Star (WD-NS) binary system. The first column shows, measurement of $\Delta t_{1}$ could be possible in the near future though still technically quite challenging. However, as is evident from the last two columns, unless the accuracy of measurement of total observed delay is raised to pico-second level and higher, which is absurd today, there is little hope to measure the effect of $\xi$.

\bigskip

\bigskip

\bigskip

\begin{center}
\textit{(b) PSR-SgrA* binary}
\end{center}

Recent observations suggest that there are probably $\sim 100$ pulsars surrounding the supermassive spinning BH SgrA* with orbital periods $\lesssim 10$ years \cite{pfahl:2004} and a few among them are expected to form PSR-BH binaries with stellar sized BH companions residing within $\sim 1$ parsec of SgrA* \cite{faucher:2011}. We shall assume the possibility that some of the pulsars cross the optical axis $OLS$ behind SgrA* making a PSR-SgrA* binary.

\begin{table}
\centering
\begin{tabular}{|c|c|c|c|c|}
\hline
$\xi /M$ & $\chi $ & $\Delta t_{1}$ ($\mu $sec) & $\Delta t_{2}$ ($\mu $sec) & $\Delta t_{3}$ ($\mu $sec) \\
\hline
$0$ & $0.1$ & $4.20$ & $1.90\times 10^{-7}$ & $9.29\times 10^{-15}$ \\
$10^{-5}$ & $0.1$ & $4.20$ & $1.71\times 10^{-7}$ & $7.53\times 10^{-15}$ \\
$10^{-4}$ & $0.1$ & $4.20$ & $9.51\times 10^{-8}$ & $2.32\times 10^{-15}$ \\
\hline
\end{tabular}
\caption{Table shows some typical RTD components $\Delta t_{1}$, $\Delta t_{2}$, $\Delta t_{3}$ calculated from Eqs.(50,53,54) for SgrA* with values given by Kato \textit{et al.} \cite{kato:2010}, viz., $M=4.2\times 10^{6}M_{\odot }$, $d_{OL}=7.6$ kpc and a \textit{unique} $a=0.44M$ so that $r_{\text{ph}}^{-}=2.15\times 10^{12} $ cm, $r_{\text{ph}}^{+}=1.51\times 10^{12}$ cm, $b=10^{7}r_{\text{ph}}^{-}$ so that $M/b<<1$. The distances $d_{LS}$, $d_{OL}$ in Fig.1 are such that the angles remain small: $\gamma_{1} \simeq \tan{\gamma_{1} = b/d_{LS}} \simeq \delta_{1} \simeq \tan{\delta_{1}}$ etc. The Kerr values are at $\xi=0$ (first row). The values of $\xi/M$ are chosen respecting the upper limit (26).}
\end{table}

In contrast to the PSR-Cygnus X1 system, we find that $\Delta t_{1}$ $\sim 4.2$ $\mu$sec allowed by the thin-lens approximation, which should be measurable with current technology provided an appropriate pulsar is identified in the future missions.

\section{Conclusions}
\label{sec:concl}
In the above, we applied our generic formula (30) for RTD to the KSBH assuming it to be a possible spinning lens partner in a binary system and investigated the effect of the string parameter $\xi$ using gravitational thin-lens approximation and PPN expansion up to third order in $\left(1/r\right)$. In our analysis, we have taken care of three important factors, often overlooked in the literature. First, the impact parameters must be $b$ and $b+a$ on the relevant sides of the lens, the latter caused by the spin $a$ or frame dragging effect. Second, the source and observer distances from the lens have been taken to be large but finite, as required by any realistic astrophysical lensing experiment. Thirdly, the null rays must pass at large distance away from the photon spheres $r_{\text{ph}}^{\pm}$ on either side of the lens. Note that photon spheres demarcate strong field limit since there is logarithmic divergence in the deflection angle so that light rays are eternally captured on the photon sphere\footnote{Note that the strong field logarithmic deflection term cannot simply be Taylor expanded to yield the weak field deflection terms. This probable misconception is clarified by Iyer and Hansen \cite{iyer:2009}. Thus for strong field deflection, one needs to adopt a completely different approach, e.g., that developed by Bozza \cite{bozza:2002}.}. Since we are interested in the weak field thin-lens limit, we considered $b=10^{n}r_{\text{ph}}^{-}$, where $n>1$ is a large number and $r_{\text{ph}}^{-}$ is always the larger of the two radii $r_{\text{ph}}^{\pm}$ such that $\frac{M}{b}<<1$, as required. And finally, we clarify that RTD is fundamentally different from the Shapiro time delay (contrary to the statements in \cite{laguna:1997}), where a single onward light ray grazing an intervening gravitating source is reflected back from a distant object in superior conjunction and the two-way travel times are added, whereas in the RTD two light rays emanate from a variable source passing behind a spinning gravitating source (lens) and the travel times are subtracted.

The method adopted in this paper can be applied with ease to any binary system assuming the source, lens and observer as aligned points. The resulting order of magnitude estimates for $\Delta t_{1}$, $\Delta t_{2}$ and $\Delta t_{3}$ from Eqs.(50,53,54) are tabulated for two illustrative binary lens systems. They are quite robust, that is, the order of magnitudes remain unaltered even when the parameters $\chi $ and $\xi $ are varied preserving small angles and the constraint (26) respectively. That's why we had presented only the two small representative Tables 1 and 2. However, these estimates are to be taken as only suggestive, since no binary system has yet been conclusively identified, although Monte Carlo simulations indicate that the number of PSR--BH binaries is expected to be significant with BH companion of a few solar masses \cite{lipunov:2005}. We have considered two binary systems, PSR-Cygnus X1 and PSR-SgrA* and the zeroth order delays have been found to be at the $\mu $sec level quite consistent with the predictions in the literature \cite{laguna:1997}. However, the higher order delays $\Delta t_{2}$ and $\Delta t_{3}$ containing the $\xi -$effect are too tiny to be measurable making the\textit{\ stringy and general relativity BHs observationally indistinguishable as of today. }This is our broad conclusion.

In detail, the results are:

$\bullet$ The analysis revealed the string induced deviations of RTD in the KSBH from those in general relativity Kerr BH ($\xi =0$). A measured non-zero value of $\xi$, which is a far cry, could have potential implications for the status of the no-hair conjecture as well as for the string theory itself.

$\bullet$ We have generalized the zeroth order Laguna-Wolszczan formula (52) for $\Delta t_{1}$ to practically realizable finite distance lensing experiments. The Eq.(52) follows when the source and observer distances from the lens are infinite and when the impact parameter is only $b$ on either side of the lens (and not $b$ and $b+a$, as required). This notwithstanding, the difference between Eqs.(50) and (52) is too minute to be measurable for large $b$ but could very well be measurable if $b$ close enough to the photon spheres.

$\bullet$ We have argued that the RTD component $\Delta t_{1}$ could be experimentally measured in terms of the observable image magnification factor determined only by $\left(\beta/\theta_{E}\right)$, where $\theta_{E}$ is the Einstein angle in the equivalent static non-aligned lens. The image at position $\theta_{-}$ on the counter-rotating side is brighter than the image at $\theta_{+}$ on the co-rotating side.

$\bullet$ The prediction of $\Delta t_{1}$ based on Eq.(50) for PSR-Cygnus X1 system is about $0.028$ $\mu$sec (Table 1). Achieving the required level of accuracy could be possible in the near future since a precision of $0.1$ $\mu$sec was achieved by \cite{vanStraten:2001} for $PSRJ0437-4715$, a bright millisecond pulsar in a White Dwarf-Neutron Star (WD-NS) binary system. However, the measurement of higher order terms $\Delta t_{2}$ and $\Delta t_{3}$ that contain $\xi$ would require better than pico-second level accuracy, which is unlikely to be attained even in the far future. As to the PSR-SgrA* binary, it is found that $\Delta t_{1}\sim 4.11$ $\mu$sec (Table 2), which should be measurable provided a suitable variable source is detected from among the pulsars orbiting SgrA* \cite{pfahl:2004,faucher:2011} and other complications are taken care of.

Even though the leading order RTD $\Delta t_{1}$[Eq.(50)] could mean a potential new test of general relativity, the strong field lens effects involving logarithmic terms could reveal surprising characteristics. To discover these, one should integrate exact null geodesics of KSBH that emerge from the variable source, pass \textit{arbitrarily} close to the photon spheres $r_{\text{ph}}^{\pm}$ on either side before reuniting at the observer.

We should mention that there are many similarities between the Kerr-Sen and Kerr-Newman black holes. In the latter, the explicit form for the post-Newtonian gravitational time delay of light signals propagating on the equatorial plane has been derived using the null geodesic \cite{uniyal:2018b}. This result is a Kerr-Newman generalization of the well known Shapiro time delay in the Schwarzschild black hole, where light is sent from Earth to a distant reflector (Mercury in superior conjunction) and the to \& from travel times of light are \textit{added} and averaged at the observer to find the net delay measured at Earth. On the other hand, the RTD has not yet been investigated in the Kerr-Newman spacetime and it would be useful to do so. Work is underway.

\section*{Acknowledgments}
The reported study was funded by RFBR according to the research Project No. 18-32-00377.

\end{document}